\documentclass[a4paper]{article}

\usepackage{arxiv}

\usepackage[utf8]{inputenc} % allow utf-8 input
\usepackage[T1]{fontenc}    % use 8-bit T1 fonts
\usepackage{hyperref}       % hyperlinks
\usepackage{url}            % simple URL typesetting
\usepackage{booktabs}       % professional-quality tables
\usepackage{amsfonts}       % blackboard math symbols
\usepackage{nicefrac}       % compact symbols for 1/2, etc.
\usepackage{microtype}      % microtypography
\usepackage{lipsum}		% Can be removed after putting your text content
\usepackage{graphicx}
\usepackage{natbib}
\usepackage{doi}
\usepackage{latexsym}
\usepackage{amsmath}
\usepackage{amssymb}
\usepackage[mathscr]{eucal}
\usepackage{graphicx,color}
\usepackage{mathtools}
\usepackage{stmaryrd}
\usepackage{url}
\usepackage{xypic}

\usepackage{bbm}
\usepackage{tikz-cd}
\usepackage{psfrag}
\def \R {{\mathbb R}}
\def \S {{\mathbb S}}
\def \L {{\mathbb L}}

\def \SO {{\mathbf{SO}}} %\mathrm
\def \SL {{\mathbf{SL}}}
\def \sl {{\mathfrak{sl}}}
\def \Ad {{\text{Ad}}}
\def \id {{\text{id}}}
\def \tr {{\text{tr}}}

\providecommand{\calF}{\mathcal{F}}
\providecommand{\mr}{\mathring}
\providecommand{\ddt}{\frac{d}{dt}}

\usepackage{amsthm}
\newtheorem{theorem}{Theorem}
\newtheorem{proposition}{Proposition}

\newtheorem{lemma}{Lemma}
\newtheorem{remark}{Remark}

\title{Nonlinear constructive observer design for direct homography estimation}%A template for the \emph{arxiv} style}

    \author{ \href{https://orcid.org/0000-0003-1116-7415}{\includegraphics[scale=0.06]{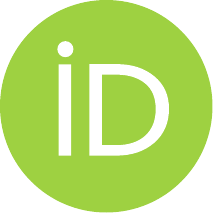}\hspace{1mm}Tarek Bouazza} \\
	I3S, CNRS, Université Côte d'Azur\\
    Sophia Antipolis, France \\
	\texttt{bouazza@i3s.unice.fr} \\
    \And
    \href{https://orcid.org/0000-0003-4391-7014}{\includegraphics[scale=0.06]{orcid.pdf}\hspace{1mm}Pieter Van Goor} \\
	Systems Theory and Robotics Group\\
    Australian National University\\
	ACT, 2601, Australia \\
	\texttt{Pieter.vanGoor@anu.edu.au} \\
	\And
    \href{https://orcid.org/0000-0002-7803-2868}{\includegraphics[scale=0.06]{orcid.pdf}\hspace{1mm}Robert Mahony} \\
	Systems Theory and Robotics Group\\
    Australian National University\\
	ACT, 2601, Australia \\
	\texttt{Robert.Mahony@anu.edu.au} \\
    \And
	\href{https://orcid.org/0000-0002-7779-1264}{\includegraphics[scale=0.06]{orcid.pdf}\hspace{1mm}Tarek Hamel} \\
    I3S, CNRS, Université Côte d'Azur\\
    Insitut Universitaire de France \\
    Sophia Antipolis, France \\
	\texttt{thamel@i3s.unice.fr} \\
}

\renewcommand{\headeright}{Author accepted version}
\renewcommand{\undertitle}{Author accepted version\thanks{\hspace{.2em}To appear in IFAC-PapersOnLine 2023 \\
\textcopyright 2023 the authors. This work has been accepted to IFAC for publication under a Creative Commons Licence CC-BY-NC-ND.}}

\renewcommand{\shorttitle}{}

\hypersetup{
pdftitle={A template for the arxiv style},
pdfsubject={q-bio.NC, q-bio.QM},
pdfauthor={David S.~Hippocampus, Elias D.~Striatum},
pdfkeywords={First keyword, Second keyword, More},
}

\begin{document}
\maketitle

\begin{abstract}
            Feature-based homography estimation approaches rely on extensive image processing for feature extraction and matching, and do not adequately account for the information provided by the image.
			Therefore, developing efficient direct techniques to extract the homography from images is essential.
			This paper presents a novel nonlinear direct homography observer that exploits the Lie group structure of $\SL(3)$ and its action on the space of image maps.
			Theoretical analysis demonstrates local asymptotic convergence of the observer.
			The observer design is also extended for partial measurements of velocity under the assumption that the unknown component is constant or slowly time-varying.
			Finally, simulation results demonstrate the performance of the proposed solutions on real images.
\end{abstract}

\section{Introduction}
	
    In projective geometry, when a planar structure is observed from multiple views, the projections of the corresponding 3D points on each image are related by a homography, which can be encoded as a matrix of unit determinant \cite{hartley2003multiple}.
	This \emph{homography matrix} is central in various applications in computer vision and robotics, such as visual tracking \cite{benhimane2007homography}, visual odometry \cite{scaramuzza2008appearance}, and visual servoing \cite{malis19992}.
	As a result, the estimation of homographies has been widely discussed and studied in the computer vision literature.

    Homography estimation methods can be broadly classified into feature-based and direct (or image-based) methods.
	The former class extracts a sparse set of salient features such as points, lines, conics, contours, etc. to perform feature matching and then compute a homography estimate \cite{kaminski2004multiple}, \cite{hua2020nonlinear}. 
	The accuracy of such estimation methods is highly dependant on the number and quality of the available features used in the estimation.
    
    In contrast, direct methods use the intensity value of all pixels of interest.
	This ensures that all the available information in the image is exploited, but generally incurs a large computational cost depending on the image resolution.
    Common direct algorithms use nonlinear optimisation methods to solve the homography estimation problem for which an initial solution is iteratively refined.
	Several minimisation algorithms have been used over the years, including Newton and Gauss-Newton methods \cite{baker2004lucas} and the Efficient Second-order Minimisation (ESM) \cite{benhimane2004real}.
    Rather than directly optimising the homography parameters as a matrix embedded in $\R^9$, alternative parametrisations have been shown to give better results.
	Examples include the four-point parametrisation \cite{baker2006parameterizing} and the $\SL(3)$ matrix exponential parametrisation \cite{mei2006homography}.
    
    In recent years, significant work has been devoted to designing nonlinear observers for feature-based estimation of homographies with dynamics.
	A number of successful approaches have exploited the $\SL(3)$ Lie group structure of the set of homographies to yield powerful stability guarantees \cite{mahony2012nonlinear}, \cite{hamel2011homography}.
	By exploiting the temporal correlations and velocity information inherent to robotics problems, these methods provide improved estimates over algorithms that simply use each image independently.
	To the authors' knowledge, this has not been investigated for direct approaches.
    
    This paper introduces a novel approach to image-based homography estimation that exploits the group action of $\SL(3)$ on the space of image maps.
	Images are modelled as square integrable functions parametrised by an evolving homography, and it is shown that there is an induced action of $\SL(3)$ on the space of such functions.
	This insight is exploited to design a nonlinear observer that estimates the unknown dynamic homography by minimising the difference between a reference image and an error image formed from the measured image and the current estimate.
	Additionally, the proposed observer is extended to estimate both homography and velocity by fusing image data with angular velocity measurements when the homography dynamics are only partially known \cite{hua2019feature}.
    
    The paper is organized into five sections, including the present introduction. Section 2 introduces the notation, homography, and some results on image maps and group actions. The image-based nonlinear observer is proposed in Section 3, and a theoretical analysis is provided to guarantee local asymptotic error convergence and derive an observer for group velocity estimation.
	Section 4 presents the results of simulations that verify the performance of the proposed observers.
	Finally, concluding remarks are presented in Section 5.
	
	%%%%%%%%%%%%%%%%%%%%%%%%%%%%%%%%%%%%%%%%%%%%%%%%%%%%%%%%%%%%%%%%%%%%%%%%%%%%%%%%
	
	%-----------------------------------------
	\section{Preliminary material}
	%-----------------------------------------

	%-----------------------------------------
	\subsection{Notation}
	%-----------------------------------------
	
	Let $e_1, e_2, e_3 \in \R^3$ denote the canonical basis, and let $I_n \in \R^{n \times n}$ be the identity matrix.
	The $2$-dimensional sphere embedded in $\R^{3}$ with radius equal to one is denoted $\mathbb{S}^2 := \{\mathbf{x} \in \R^{3} \mid \vert \mathbf{x} \vert = 1\}$. 
	Given $a \in \R^3$, let $a_\times$ denote the skew symmetric matrix associated to the cross
	product,
	\begin{align*}
		a_\times := \begin{pmatrix}
			0    & -a_3 & a_2  \\
			a_3  & 0    & -a_1 \\
			-a_2 & a_1  & 0
		\end{pmatrix}.
	\end{align*}
	Then $a_\times b = a \times b$ for all $ a, b \in \R^3$. 
	The sphere projection is defined $\Pi_{\S^2}(v) :=  v / |v|$ for all $v \in \R^3 /\{0\}$.
	The projection onto the tangent space of the unit sphere at the point $\mathbf{x} \in \mathbb{S}^2$ is denoted $ \Pi_{\mathbf{x}} := (I_{3} - \mathbf{x} \mathbf{x}^\top) $.
	
	For any two matrices $A, B \in \R^{3\times 3}$ the Euclidean matrix inner product and Frobenius norm are defined as
	\begin{align}
		\langle A, B \rangle &:= \tr (A^\top B), &
		\vert A \vert &:= \sqrt{\langle A, A \rangle},
		\label{eq:matrix_inner_product}
	\end{align}
	respectively.
	
	Given a measureable space $X$, such as a suitable subset $X \subset \R^3$, a square integrable function is a map $f: X \to \R$ such that
	\begin{align*}
		\Vert f \Vert^2 := \int_X \vert f(\mathbf{x}) \vert^2 d \mathbf{x},
	\end{align*}
	exists and is finite.
	The space of all square integrable functions on $X$ is denoted $\L^2(X)$, and is a Hilbert space; that is, a real vector space equipped with an inner product defined by
	\begin{align*}
		\langle f, g \rangle := \int_X \vert f(\mathbf{x}) g(\mathbf{x}) \vert d \mathbf{x},
	\end{align*}
	for all $f,g \in L^2(X)$.
	
    The special orthogonal group is denoted $\SO(3)$.
	\begin{align*}
	    \SO(3) := \{ R \in \R^{3 \times 3} : \det(R) = 1, RR^\top = R^\top R = I_3 \}.
	\end{align*}
	%-----------------------------------------
	\subsection{Special Linear Group $\SL(3)$}
	%-----------------------------------------
	
	The special linear group $\SL(3)$ is a matrix Lie group with matrix Lie algebra $\mathfrak{sl}(3)$, defined by
	\begin{align*}
		\SL(3) &:= \{ H \in \R^{3 \times 3} : \det(H) = 1 \}, \\
		\mathfrak{sl}(3) &:= \{ U \in \R^{3\times 3} :\; \text{tr}(U) = 0 \},
	\end{align*}
	respectively.
	The exponential map $\exp: \sl(3) \to \SL(3)$ defines a local diffeomorphism from a neighbourhood of $0 \in \sl(3)$ to a neighbourhood of $I_3 \in \SL(3)$,
	The adjoint operator $\Ad : \SL(3) \times \mathfrak{sl}(3) \to \mathfrak{sl}(3)$ is defined by
	\begin{align*}
		\Ad_H U := HUH^{-1},
	\end{align*}
	for all $H \in \SL(3)$ and  $U \in \mathfrak{sl}(3)$.

	The unique orthogonal projection of $\R^{3\times 3}$ onto $\sl(3)$, with respect to the inner product \eqref{eq:matrix_inner_product}, is given by
	\begin{align*}
		\mathbb{P}_{\sl(3)}(M) := M - \frac{\tr(M)}{3}I_3 \, \in \sl(3)
	\end{align*} 
	for all $M \in \R^{3\times 3}$.
	
	%-----------------------------------------
	\subsection{Definition of Homography}
	%-----------------------------------------
	
	Consider a camera moving in space observing a planar scene. Let $\{ \mathring{\mathcal{C}} \}$ and $\{ \mathcal{C}_t \}$ denote the reference frame and the current camera frame, respectively. 
	Let $R \in \SO(3)$ and $\xi \in \R^3$ denote the orientation and translation, respectively, of the camera frame $\{ \mathcal{C}_t \}$ at time $t$ with respect to the reference frame $\{ \mathring{\mathcal{C}} \}$.
	
	Let $d$  denote  the  distance  from  the origin of $\{ \mathcal{C}_t \}$ to the planar scene and $\eta \in \mathbb{S}^2$  the normal vector  pointing  to  the  scene  expressed  in $\{ \mathcal{C}_t \}$.
	Then the planar scene is parametrized by 
	\begin{align*}
		\mathbf{\Pi} := \{  \mathcal{P} \in \R^3 \mid \eta^\top \mathcal{P} - d = 0 \}.
	\end{align*}
	Now, let $\mathring{\mathcal{P}} \in \{ \mathring{\mathcal{C}} \}$ and $\mathcal{P} \in \{ \mathcal{C}_t \}$ be the $3D$ coordinate vectors of a point belonging to the planar scene, related by
	\begin{equation} \label{eqn:mp}
		\mathring{\mathcal{P}} = R \mathcal{P} + \xi%P = R^\top (\mathring{P} + \xi)
	\end{equation}
	Then applying the plane constraint $\frac{\eta^\top \mathcal{P}}{d} = 1$  yields
	\begin{equation} \label{eqn:phe}
		\mathring{\mathcal{P}} = \left( R + \frac{\xi \eta^\top}{d}\right) \mathcal{P}.
	\end{equation}
	By projecting the coordinates onto the unit sphere as
	\begin{equation} \label{eqn:pc}
		\mathring{p} := \Pi_{\S^2}(\mathring{\mathcal{P}}), \;\;  p := \Pi_{\S^2}(\mathcal{P}),
	\end{equation}
	the projected points satisfy\footnote{$\cong$ denotes equality up to a multiplicative constant.}
	\begin{equation} \label{eqn:hce}
		\mathring{p}  \cong \left( R + \frac{\xi \eta^\top}{d} \right) p \cong H p,
	\end{equation}
	where the projective mapping $H:= R + \frac{\xi \eta^\top}{d}$ is the Euclidean homography that maps Euclidean coordinates of the scene's points from $\{ \mathcal{C}_t \}$ to $\{ \mathring{\mathcal{C}} \}$.
	
	The homography matrix $H$ is only defined up to a scale factor, and so every homography can be associated with a unique matrix $\bar{H} \in \SL(3)$ by rescaling
	\begin{equation}
		\bar{H} := \det(H)^{-\frac{1}{3}} H,
	\end{equation}
	such that $\det{\bar{H}} = 1$ and thus $\bar{H} \in \SL(3)$.
	For the remainder of this paper all homographies $H$ are taken to be appropriately scaled so that $H \in \SL(3)$.
	
	\subsection{Images as Maps on the Sphere}
	
	Consider an image map $\mathcal{I} : \mathcal{U} \subset \R^2 \to [0,1] \subset \R$, where $\mathcal{U}$ is the image domain.
	Let $h: \mathcal{S} \subset \S^2 \to \mathcal{U}$ be the smooth bijective transformation between ray directions on the unit sphere and image pixel coordinates, and define $I : \mathcal{S} \to [0,1] \subset \R$ by $I(\mathbf{x}) = \mathcal{I}(h(\mathbf{x}))$; that is, the diagram
	\[
	\xymatrix{
		\mathcal{S} \ar[r]^h \ar[rd]^I &  \mathcal{U} \ar[d]^{\mathcal{I}} \\
		& [0,1] 
	}
	\]
	commutes.
	We will use this formulation in the remainder of the paper, as working on the sphere adds to the clarity of the mathematical presentation.
	
	\medbreak
	\begin{remark}
		%    \todo{[Move this to earlier section on $h$.]}
		In the perspective camera model, when the intrinsic camera parameters are known, the projection $h$ is given by
		$$
		h(\mathbf{x}) := \begin{pmatrix}
			f_u & 0 & u_0 \\ 0 & f_v & v_0 
		\end{pmatrix} \frac{\mathbf{x}}{e_3^\top \mathbf{x}},
		$$
		where $(f_u, f_v)$ and $(u_0, v_0)$ represent the focal length in pixels and the optical center coordinates, respectively, with $h(\mathbf{x}) = (u, v)^\top$ denoting the projected pixel coordinates in the image plane. 
		
		The derivative $DI(\mathbf{x})$ of the image map $I$ at $\mathbf{x}$ can be expressed as follows %yields
		\begin{align*}
			DI(\mathbf{x}) &= D \mathcal{I}(u, v) D h(\mathbf{x}), \\ 
			&= D \mathcal{I}(u, v)
			\begin{pmatrix}
				f_u & 0 & u_0 \\ 0 & f_v & v_0 
			\end{pmatrix} \frac{1}{e_3^\top \mathbf{x}}\left( I_3 - \frac{\mathbf{x}e_3^\top}{|\mathbf{x}e_3^\top|}\right),
			%\frac{e_3^\top \mathbf{x} I_3 - \mathbf{x} e_3^\top}{(e_3^\top \mathbf{x})^2} %|\mathbf{X}| K
		\end{align*}
		where $D \mathcal{I}(u,v) = \left. \begin{bmatrix} D_u \mathcal{I}, D_v \mathcal{I} \end{bmatrix} \right|_{(u,v)}$ denotes the transpose of the image $\mathcal{I}$ gradient at $(u, v)^\top$.
	\end{remark}

	\medbreak
	\begin{proposition}[\cite{hua2020nonlinear}] \label{prop1}
		The function $\phi: \SL(3) \times \S^2 \mapsto \S^2$ given by
		\begin{equation*}
			\phi(H, \mathbf{x}) := \frac{H^{-1} \mathbf{x}}{|H^{-1} \mathbf{x}|},
		\end{equation*}
		is a right group action of $\SL(3)$ on $\S^2$. 
	\end{proposition}
	The right action $\phi$ on $\S^2$ can be seen as a coordinate transformation referred to as warping \cite{mei2008efficient}. 
	
	In the remaining of this work image maps are assumed to be elements of $\L^2(\S^2)$.
	The action $\phi$ can be used to construct a left action $\Phi$ on image maps. 
	\medbreak
	\begin{lemma} \label{lemma1}
		Let $\L^2(\S^2)$ denote the set of square integrable functions $\S^2 \to \R$.
		Then $\Phi: \SL(3) \times \L^2(\S^2) \to \L^2(\S^2)$, defined by 
		\begin{equation*}
			\Phi(H,I) := I \circ \phi_{H^{-1}},
		\end{equation*}
		is a right group action.
	\end{lemma}
	\begin{proof}
		Checking the identity property, for any $I \in \L^2(\S^2)$ and any $\mathbf{x} \in \S^2$,
		\begin{align*}
			\Phi(\id, I)(\mathbf{x}) &= I \circ \phi_{\id}(\mathbf{x}) = I(\mathbf{x}).
		\end{align*}
		Now checking the composition property, for any $I \in \L^2(\S^2)$, any $\mathbf{x} \in \S^2$, and any $G, H \in \mathbf{SL}(3)$, one has
		\begin{align*}
			\Phi(G, \Phi(H, I)) &= \Phi(G, I \circ \phi_{H^{-1}}), \\
			&= f \circ \phi_{H^{-1}} \circ \phi_{G^{-1}}, \\
			&= f \circ \phi_{G^{-1} H^{-1}}, \\
			&= f \circ \phi_{(HG)^{-1}} = \Phi(HG, I).
		\end{align*}
		Finally, checking that $\L^2(\S^2)$ is closed under $\Phi$, observe that, for any $I \in \L^2(\S^2)$ and $H \in \SL(3)$,
		\begin{align*}
			\Vert \Phi(H,I) \Vert^2
			&= \int_{\S^2}  \Phi(H,I)(\mathbf{x} )^2 d\mathbf{x}, \\
			&= \int_{\S^2}  I( \phi(H^{-1},\mathbf{x}) )^2 d\mathbf{x}, \\
			&= \int_{\phi_{H^{-1}}(\S^2)}  I(\mathbf{y})^2 \vert \det(D \phi_{H})(\mathbf{y}) \vert d\mathbf{y}, \\
			&= \int_{\S^2}  I(\mathbf{y})^2 d\mathbf{y} = \Vert I \Vert^2, 
		\end{align*} %, \\ &
		where the second last step follows from the fact that $\phi_{H^{-1}}(\S^2) = \S^2$ for any $H \in \SL(3)$, and that $\det(D \phi_{H})(\mathbf{y}) = \det(\Pi_{\phi_{H}(\mathbf{y})} H^{-1}) = \det(H^{-1}) = 1$.
	\end{proof}
	
	Lemma \ref{lemma1} shows how the action $\phi$ of $\SL(3)$ on $\S^2$ induces an action $\Phi$ of $\SL(3)$ on $\L^2(\S^2)$ and consequently that homographies act on image maps of planar scenes.
	
	\section{Observer design}
	Consider a camera that is moving according to some measured linear and angular velocity while capturing images of a planar scene.
	Let $H(t) \in \SL(3)$ denote the homography that maps the current image to a fixed reference image, with kinematics given by 
	\begin{equation} \label{eq:sys}
		\dot{H} = HU, 
	\end{equation}
	where the group velocity $U \in \sl(3)$ induced by the relative motion between the camera and the planar scene satisfies [\cite{mahony2012nonlinear}, Lem. 5.3]
	\begin{equation} \label{eq:groupvel}
		U = \Omega_{\times} + \frac{V \eta^\top}{d} - \frac{\eta^\top V}{3d}I_3, 
	\end{equation}
	with $\Omega \in \R^3$ and $V \in \R^3$ denoting the angular and linear velocities of the camera, expressed in $\{\mathcal{C}_t\}$, both assumed to be bounded.
	
	\subsection{Homography Observer Design}
	Let $\hat{H} \in \SL(3)$ denote the estimate of $H$. Assume that $U \in \sl(3)$ is available, and define the dynamics of the proposed observer to be
	\begin{align} \label{eq:observer}
		\dot{\hat{H}} &= \hat{H}U + \Delta \hat{H}, &
		\hat{H}(0) &= I_3,
	\end{align}
	where $\Delta \in \sl(3)$ is the correction term that remains to be designed.
	
	Define the group error $E := \hat{H}H^{-1}$.
	Then the dynamics of $E$ are
	\begin{equation} \label{eq:error}
		\begin{split}
			\dot{E} &=  \hat{H}UH^{-1} + \Delta H\hat{H}^{-1} - \hat{H}UH^{-1}
			= \Delta E
		\end{split}
	\end{equation}
	The problem of the observer design is to identify a correction term $\Delta \in \sl(3)$ that ensures that the group error $E$ converges to the identify $I_3$, and therefore
	\begin{equation*}
		\hat{H} = \hat{H} H^{-1} H = E H \to I_3 H = H.
	\end{equation*}
	
	Let $\mathring{\mathcal{I}}$ be the reference image.
	We denote $\mathcal{R}_{\mr{\mathcal{I}}} \subseteq \mathcal{U}$ a region of size $(h \times w)$ of $\mathring{\mathcal{I}}$ corresponding to the projection of the planar region of the scene on the reference image plane, and $\mathcal{X} \subseteq \mathcal{S}$ the projection of $\mathcal{R}_{\mathring{\mathcal{I}}}$ onto the unit sphere.
	
	Define $\mathring{I}: \mathcal{X} \rightarrow [0, 1]$ to be the reference image map. Then, the current image map is given by 
	\begin{align*}
		I &:= \Phi(H, \mathring{I}), \\
		I(\mathbf{x}) &= \Phi(H, \mathring{I})(\mathbf{x})
		= \mathring{I}(\phi(H^{-1},\mathbf{x})).
	\end{align*}
	The \emph{warped image map} is defined by exploiting the group action property,
	\begin{equation}
		I^e := I \circ \phi_{\hat{H}} = \Phi(E^{-1}, \mathring{I}).
	\end{equation}
	
	Both $\mathring{I}$ and $I$ are assumed to be $\L^2(\mathcal{X})$.
	Define the cost function $\mathcal{F}: \SL(3) \mapsto \R^+$,
	\begin{align}
		\mathcal{F}(E) &= \frac{1}{2} \Vert I^e - \mathring{I} \Vert^2, \notag \\ &= \frac{1}{2} \int_{\mathcal{X}} \left( \Phi(E^{-1}, \mathring{I})(\mathbf{x}) - \mathring{I}(\mathbf{x}) \right)^2 d \mathbf{x}.
		\label{eq:cost} %, \notag \\ &
	\end{align}
	This function is differentiable as long as the image maps $I$ and $\mathring{I}$ are.
	\begin{lemma}
		Suppose that the image maps $\mathring{I}$ and $I$ are differentiable (and therefore continuous) and let $\text{grad} \mathring{I}(\mathbf{x})$ denote the reference image map gradient at $\mathbf{x}$. If
		%$\forall \Delta \in \sl(3), \; \exists \mathbf{x} \in \mathcal{X}$ such that
		\begin{align}
			\mathrm{Span}\{ \text{grad} \mathring{I}(\mathbf{x}) \mathbf{x}^\top \; | \; \forall \mathbf{x} \in \mathcal{X}\} = \sl(3),
			\label{eq:span_of_gradients}
		\end{align}
		then the cost function \eqref{eq:cost} has an isolated global minimum at $E = I_3$.
	\end{lemma}
	
	\begin{proof}
		It is straightforward to see that $\calF(I_3) = 0$ and is therefore a global minimum.
		To show that this is an isolated minimum, it suffices to show that the Hessian of $\calF$ at $I_3$ is non-degenerate.
		The first order derivative of $\mathcal{F}$ is computed as follows,
		\begin{align*}
			D\mathcal{F}(E)[\Delta E] 
			&= \int_{\mathcal{X}} \left(I^e(\mathbf{x}) - \mathring{I}(\mathbf{x})\right) D(\mathring{I} \circ \phi_{\mathbf{x}})(E)[\Delta E]d\mathbf{x}, \\
			&= \int_{\mathcal{X}} \left. \left(I^e(\mathbf{x}) - \mathring{I}(\mathbf{x})\right) \frac{d}{dt} \mathring{I} \circ \phi(\exp(t\Delta)E, {\mathbf{x}}) \right|_{t=0}d\mathbf{x}, \\
			&= \int_{\mathcal{X}} \left. \left(I^e(\mathbf{x}) - \mathring{I}(\mathbf{x})\right) \frac{d}{dt} \mathring{I} \circ \phi\left(E, \phi(\exp(t\Delta), \mathbf{x})\right) \right|_{t=0}d\mathbf{x}, \\
			&= \int_{\mathcal{X}} \left. \left(I^e(\mathbf{x}) - \mathring{I}(\mathbf{x})\right) \frac{d}{dt} I^e \circ \phi(\exp(t\Delta), \mathbf{x}) \right|_{t=0}d\mathbf{x}, \\
			&= \int_{\mathcal{X}} \left(I^e(\mathbf{x}) - \mathring{I}(\mathbf{x})\right) D I^e(\mathbf{x}) D \phi_\mathbf{x}(I_3)[\Delta] d\mathbf{x}. 
		\end{align*}
		where $DI^e(\mathbf{x})$ denotes the transpose of the warped image gradient at $\mathbf{x}$; that is, $DI^e(\mathbf{x}) = \mathrm{grad}I^e(\mathbf{x})^\top$. 
		
		The differential $D \phi_\mathbf{x}(I_3)[\Delta] $ is given by
		\begin{align*}
			D \phi_\mathbf{x}(I_3)[\Delta] &= \left. \frac{d}{dt} \right|_{t=0} \phi((\exp(t\Delta), \mathbf{x}), \\
			&= (\mathbf{x} \mathbf{x}^\top - I_3) \Delta \mathbf{x} = - \Pi_{\mathbf{x}} \Delta \mathbf{x},
		\end{align*}
		and, hence
		\begin{align*}
			D\mathcal{F}(E)[\Delta E] &= - \int_{\mathcal{X}} \left( I^e(\mathbf{x}) - \mathring{I}(\mathbf{x}) \right) DI^e(\mathbf{x})\Pi_{\mathbf{x}} \Delta \mathbf{x} d\mathbf{x}.
		\end{align*}
		The second order derivative of $\mathcal{F}$ about $I_3$ is computed by
		\begin{align*}
		    D^2\mathcal{F}(I_3)[\Delta, \Delta] 
		    &= - \left. \int_{\mathcal{X}} \left( D(\mathring{I} \circ \phi_{\mathbf{x}})(E)[\Delta E] \right) DI^e(\mathbf{x})\Pi_{\mathbf{x}} \Delta \mathbf{x} d\mathbf{x}\right|_{E=I_3}, \\ 
			&= - \int_{\mathcal{X}} \left( D \mathring{I}(\mathbf{x}) D \phi_\mathbf{x}(I_3)[\Delta] \right) D\mathring{I}(\mathbf{x})\Pi_{\mathbf{x}} \Delta \mathbf{x} d\mathbf{x}, \\ 
			&= \int_{\mathcal{X}} \left( D\mathring{I}(\mathbf{x})\Pi_{\mathbf{x}} \Delta \mathbf{x} \right)^2 d\mathbf{x},
		\end{align*}
		and since $ D\mathring{I}(\mathbf{x})\Pi_\mathbf{x} = D\mathring{I}(\mathbf{x})$,
		\begin{align}
			D^2\mathcal{F}(I_3)[\Delta, \Delta]
			&= \int_{\mathcal{X}} \left( D\mathring{I}(\mathbf{x}) \Delta \mathbf{x} \right)^2 d\mathbf{x}, \notag \\
			&= \int_{\mathcal{X}} \left( \tr( D\mathring{I}(\mathbf{x}) \Delta \mathbf{x}) \right)^2 d\mathbf{x}, \notag \\
			&= \int_{\mathcal{X}} \left( \left\langle D\mathring{I}(\mathbf{x})^\top \mathbf{x}^\top, \Delta \right\rangle \right)^2 d\mathbf{x}, \notag \\
			&= \int_{\mathcal{X}} \left\langle \mathrm{grad}\mathring{I}(\mathbf{x}) \mathbf{x}^\top, \Delta \right\rangle^2 d\mathbf{x}.
			\label{eq:hessian_equation}
		\end{align}
		Observe that $\text{grad}\mathring{I}(\mathbf{x})\mathbf{x}^\top \in \sl(3)$ as
		\begin{align*}
			\tr(\text{grad}\mathring{I}(\mathbf{x})\mathbf{x}^\top) = \tr(D\mathring{I}(\mathbf{x})\mathbf{x}) = 0.
		\end{align*}
		
		Now suppose the Hessian is degenerate; i.e. there exists $\Delta \in \sl(3)$ such that $D^2\mathcal{F}(I_3)[\Delta, \Delta] = 0$.
		Then, since the integrand in \eqref{eq:hessian_equation} is non-negative for every $\mathbf{x}$ and $\mr{I}$ is continuous, it must be that 
		\begin{align*}
			\left\langle \mathrm{grad}\mathring{I}(\mathbf{x}) \mathbf{x}^\top, \Delta \right\rangle = 0,
		\end{align*}
		for all $\mathbf{x} \in \mathcal{X}$.
		This contradicts the assumption \eqref{eq:span_of_gradients}, and therefore the Hessian \eqref{eq:hessian_equation} must be non-degenerate.
	\end{proof}
	
	\medbreak
	\begin{theorem}
		Consider the homography observer (\ref{eq:observer}) with correction term $ \Delta \in \sl(3)$ given by
		\begin{equation} \label{eq:deltaconst}
			\Delta = k_{\Delta}  \int_{\mathcal{X}} \Tilde{I}(\mathbf{x})  \mathrm{grad}I^e(\mathbf{x}) \mathbf{x}^\top d\mathbf{x}, 
		\end{equation}
		where $\Tilde{I}(\mathbf{x}) = I^e(\mathbf{x}) - \mathring{I}(\mathbf{x})$ denotes the image intensity difference at $\mathbf{x}$ and $k_{\Delta} > 0$ a chosen gain.
		Then the equilibrium $E = I_3$ of the autonomous system \eqref{eq:error} is locally asymptotically stable. %\todo{[exponentially?]} 
	\end{theorem}
	
	\begin{proof}
		The time derivative of $\mathcal{F}$ is given by 
		\begin{align*}
			%\dot{f}_\mathbf{x} 
			\dot{\mathcal{F}}(E) %&= \frac{d}{dt} \langle I^e - \mathring{I}, I^e - \mathring{I}  \rangle_{\L_2} \\
			&=\int_{\mathcal{X}} \left(I^e(\mathbf{x}) - \mathring{I}(\mathbf{x})\right) \frac{d}{dt} \Phi(E^{-1}, \mathring{I})(\mathbf{x}) d\mathbf{x}, \\
			&= \int_{\mathcal{X}} \left(I^e(\mathbf{x}) - \mathring{I}(\mathbf{x})\right) \frac{d}{dt} \mathring{I} \circ \phi_{\mathbf{x}}(E)d\mathbf{x}, \\
			&= \int_{\mathcal{X}} \left(I^e(\mathbf{x}) - \mathring{I}(\mathbf{x})\right) D(\mathring{I} \circ \phi_{\mathbf{x}})(E)[\Delta E]d\mathbf{x}, \\
			&= \int_{\mathcal{X}} \left(I^e(\mathbf{x}) - \mathring{I}(\mathbf{x})\right) D I^e (\mathbf{x}) D \phi_\mathbf{x}(I_3)[\Delta] d\mathbf{x}.
		\end{align*}
		It follows that 
		\begin{align*}
			\dot{\mathcal{F}}(E)
			&= - \int_{\mathcal{X}} \Tilde{I}(\mathbf{x}) DI^e(\mathbf{x}) \Pi_{\mathbf{x}} \Delta \mathbf{x} d\mathbf{x}, \\
			&= - \int_{\mathcal{X}} \langle \Tilde{I}(\mathbf{x}) \mathrm{grad}I^e(\mathbf{x}),  \Delta \mathbf{x} \rangle d\mathbf{x}, \\
			&= - \langle  \int_{\mathcal{X}}\Tilde{I}(\mathbf{x}) \mathrm{grad}I^e(\mathbf{x}) \mathbf{x}^\top d\mathbf{x} , \Delta \rangle. 
		\end{align*}
		Then, by choosing $\Delta$ as in \eqref{eq:deltaconst}, one obtains
		\begin{align*}
			\dot{\mathcal{F}}(E) = - \frac{1}{k_{\Delta}} |\Delta|^2 \leq 0. %= - |\Delta|^2 - k 
		\end{align*}
		The time derivative $\dot{\mathcal{F}}$ is negative semi-definite, and equal to zero when $\Delta = 0$.
		This implies, by application of LaSalle's theorem, that $\Delta$ converges asymptotically to zero. Then from \eqref{eq:error}, one concludes that $E = I_3$ is locally asymptotically stable.
	\end{proof}
 
\subsection{Observer design with partially known velocity}
The group velocity $U$, as defined by (\ref{eq:groupvel}), depends on the variables $d$ and $\eta$ that are unknown and cannot be directly extracted from measurements.
In \cite{hua2019feature} the authors showed that $U$ can be decomposed as follows 
%According to [cite] it can be expressed as follows when 
\begin{equation}
	U = \Omega_\times + \Gamma, \;\; \text{with} \; \Gamma = \frac{V\eta^\top}{d} - \frac{\eta^\top V}{3d}I_3,
\end{equation}
where the angular velocity $\Omega$ is the measurable part, which can be obtained from the gyroscope, and $\Gamma$ represents the non-measurable part that has to be estimated.

Proceeding as in \cite{hua2019feature} and assuming that $\ddt \xi/d$ is constant or slowly time-varying (the situation when the camera is moving at a constant velocity parallel to the scene or converging exponentially towards it), and using the fact that $V = R^\top \dot{\xi}$ and $\dot{\eta} = \eta \times \Omega$, yields %one easily verifies that 
$$\dot{\Gamma} = \left[ \Gamma, \Omega_\times\right],$$
with $\left[ \Gamma, \Omega_\times\right] = \Gamma \Omega_\times - \Omega_\times \Gamma$ is the Lie bracket.

The observer takes the following form
\begin{equation} \label{eq:observervel}
	\begin{split}
		&\dot{\hat{H}} = \hat{H}(\Omega_\times + \hat{\Gamma}) + \Delta\hat{H}, \\
		&\dot{\hat{\Gamma}} = [ \hat{\Gamma}, \Omega_\times ] + k_\Gamma \Ad_{\hat{H}^\top}\Delta,
	\end{split}
\end{equation}
with $k_\Gamma >0$.
Define the velocity error $\tilde{\Gamma} = \Gamma - \hat{\Gamma}$, then the error kinematics are given by
\begin{equation} \label{eq:observererror}
	\begin{split}
		\dot{E} &= (\Delta - \Ad_{\hat{H}}\tilde{\Gamma})E, \\
		\dot{\tilde{\Gamma}} &= [\tilde{\Gamma}, \Omega_\times] - k_\Gamma \Ad_{\hat{H}^\top}\Delta.
	\end{split}
\end{equation}
\medbreak
\begin{proposition}
	Consider the proposed observer \eqref{eq:observervel} with the correction $\Delta$ given by \eqref{eq:deltaconst}.
	Then the equilibrium $(E, \tilde{\Gamma}) = (I_3, 0)$ is locally asymptotically stable.
\end{proposition}

\begin{proof}
	Consider the following cost function 
	\begin{align*}
		\mathcal{F}^\star(E, \tilde{\Gamma}) 
		&= \frac{1}{2} \int_{\mathcal{X}} \left( \Phi(E^{-1}, \mathring{I}) (\mathbf{x}) - \mathring{I}(\mathbf{x}) \right)^2 d\mathbf{x} + \frac{1}{2 k_{\Delta} k_\Gamma} | \tilde{\Gamma} |^2.
	\end{align*}
	Differentiating $\Phi(E^{-1}, \mathring{I})(\mathbf{x})$ yields
	\begin{align*}
		\frac{d}{dt} \Phi(E^{-1}, \mathring{I})(\mathbf{x}) &= D I^e(\mathbf{x}) D \phi_\mathbf{x}(E)\left[\left(\Delta - \Ad_{\hat{H}}\tilde{\Gamma}\right)E \right], \\
		&= D I^e(\mathbf{x}) D \phi_\mathbf{x}(I_3)[\Delta - \Ad_{\hat{H}}\tilde{\Gamma} ], \\
		&= - DI^e(\mathbf{x}) \Pi_\mathbf{x} \left( \Delta - \Ad_{\hat{H}}\tilde{\Gamma} \right) \mathbf{x}
	\end{align*}
	Using the fact that $\langle [\tilde{\Gamma}, \Omega_\times], \tilde{\Gamma} \rangle = 0$,
	it follows that the time derivative of $\mathcal{F}^\star$ is 
	\begin{align*}
		\dot{\mathcal{F}}^\star(E, \tilde{\Gamma}) &= \int_{\mathcal{X}} \tilde{I}(\mathbf{x}) \frac{d}{dt} \Phi(E^{-1}, \mathring{I})(\mathbf{x}) d\mathbf{x} - \langle \frac{1}{k_{\Delta}} \Ad_{\hat{H}^\top} \Delta, \tilde{\Gamma} \rangle, \\ 
		&=  - \int_{\mathcal{X}} \Tilde{I}(\mathbf{x})  DI^e(\mathbf{x}) \left( \Delta - \Ad_{\hat{H}}\tilde{\Gamma} \right)\mathbf{x} d\mathbf{x} - \langle \frac{1}{k_{\Delta}} \Ad_{\hat{H}^\top} \Delta, \tilde{\Gamma} \rangle, \\ 
		&=\begin{multlined}[t][7cm]
			- \langle \int_{\mathcal{X}}\Tilde{I}(\mathbf{x}) \mathrm{grad}I^e(\mathbf{x})\mathbf{x}^\top d\mathbf{x}, \Delta \rangle \\
			- \langle \Ad_{\hat{H}^\top} \left( \frac{1}{k_{\Delta}} \Delta - \int_{\mathcal{X}}\Tilde{I}(\mathbf{x}) \mathrm{grad}I^e(\mathbf{x})\mathbf{x}^\top d\mathbf{x} \right), \tilde{\Gamma} \rangle .
		\end{multlined}
	\end{align*}
	Lastly, by inserting the expression \eqref{eq:deltaconst} of $\Delta$, we obtain 
	$$\dot{\mathcal{F}}^\star(E, \tilde{\Gamma}) = - \frac{1}{k_{\Delta}} | \Delta |^2 \leq 0.$$
	The time derivative $\dot{\mathcal{F}}^\star$ is negative semi-definite, and equal to zero when $\Delta = 0$.
	Given that $\Omega$ is bounded, it follows that $\Delta$ converges asymptotically to zero using Barbalat's lemma. Accordingly, the left-hand side of the cost $\mathcal{F}^\star$ converges to zero and $| \tilde{\Gamma} |^2$ converges to a constant.
	
    Since $E \rightarrow I_3$ and $\Delta \rightarrow 0$, it follows from \eqref{eq:observererror} that $ \lim_{t \rightarrow + \infty} \dot{E} = - \Ad_{\hat{H}} \tilde{\Gamma} = 0$. 
    One then concludes that $(E, \tilde{\Gamma}) = (I_3, 0)$ is locally asymptotically stable.
\end{proof}
%}

\begin{remark}
    If $V/d$ remains constant or slowly time-varying (the situation where the camera follows a circular trajectory on the scene), then the same outcome can be obtained using the same proof. In this case, we have
	$$U = \Omega_\times + \mathbb{P}_{\sl(3)}(\Gamma_1), \quad \Gamma_1 = \frac{V\eta^\top}{d}, $$
	with  
	$\dot{\Gamma}_1 = \Gamma_1\Omega_\times $, and the observer has the following form
	\begin{equation}
		\begin{split}
			&\dot{\hat{H}} = \hat{H}\left(\Omega_\times +\mathbb{P}_{\sl(3)}(\hat{\Gamma}_1)\right) + \Delta\hat{H}, \\
			&\dot{\hat{\Gamma}}_1 = \hat{\Gamma}_1\Omega_\times + k_{\Gamma_1} \Ad_{\hat{H}^\top}\Delta.
		\end{split}
	\end{equation}
\end{remark}

%-----------------------------------------------
\section{Simulation results}
%-----------------------------------------------

This section presents the simulation experiment conducted to verify the performance of the observers \eqref{eq:observer} and \eqref{eq:observervel}.
From a reference image of a resolution of $256 \times 254$ pixels, the image at a given time was generated using $I = \Phi(H, \mr{I})$.

The true homography dynamics were defined according to \eqref{eq:sys}, with initial condition and group velocity
\begin{align*}
    H(0) &= \begin{pmatrix} 1.0308 & 0.0507 &  0.0867 \\
	-0.051 & 1.0309 & -0.144 \\
	0 & 0 &  0.9388 \end{pmatrix}, & 
	U(t) = \begin{pmatrix} 0 & 0 &  -0.1 \\
	0 & 0 & 0.1 \\
	0 & 0 &  0 \end{pmatrix},
\end{align*}
%and 
corresponding to a constant translational motion parallel to the scene, with $\Omega \equiv 0$ and $\ddt \xi/d = (-0.1, 0.1, 0)^\top$.

In the first simulation, $U$ was assumed to be available. The observer dynamics were defined according to \eqref{eq:observer} with initial condition $\hat{H}(0) = I_3 $ 
and gain $k_\Delta = 0.1$. 

In the second simulation, $U$ was only partially known and the dynamics were defined by \eqref{eq:observervel} with initial conditions $\hat{H}(0) = I_3$, $\hat{\Gamma}(0) = 0_{3 \times 3}$ and gains $k_\Delta = 0.1$, $k_\Gamma = 2$.

Both simulations used Euler integration for 3~s with a time-step of 0.02 seconds. The homography dynamics were integrated using the matrix exponential map to ensure that they remained in the Lie group $\SL(3)$ for all time.

\begin{figure}[h]
\centering
\includegraphics[scale=.35]{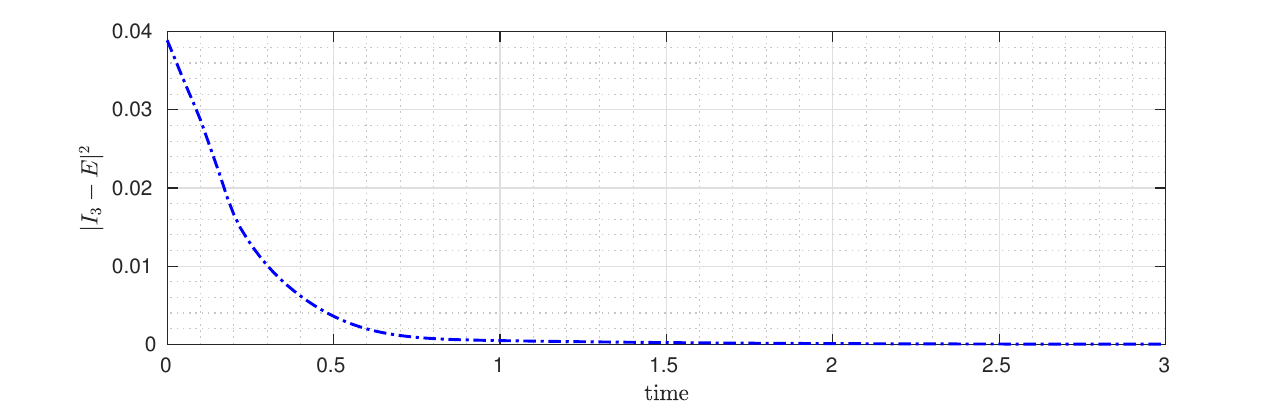} \\
\includegraphics[scale=.35]{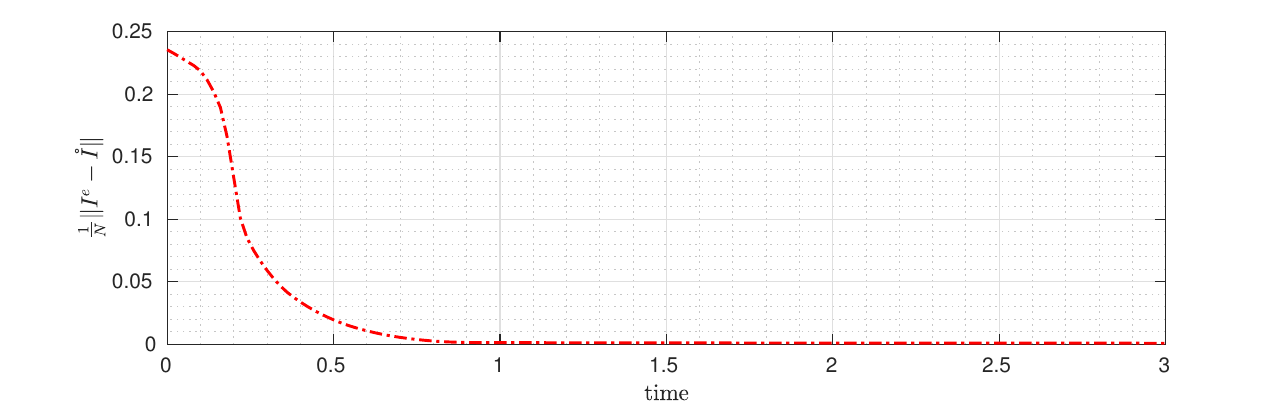}
\caption{Time evolution of the homography estimation error (blue) and image intensity error divided by the number of pixels (red) of the first simulation. }
\label{fig:errors}
\end{figure}

\begin{figure}[h]
\centering
\includegraphics[width=8.3cm,height=2.1cm]{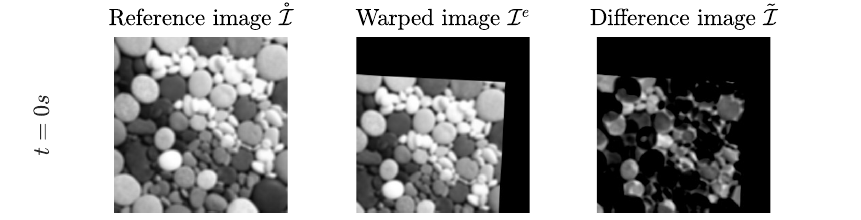}
\includegraphics[width=8.25cm,height=1.9cm]{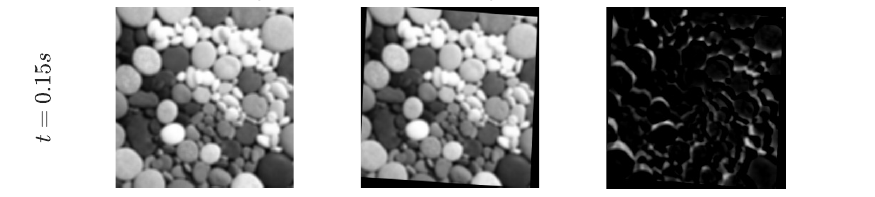}
\includegraphics[width=8.3cm,height=1.9cm]{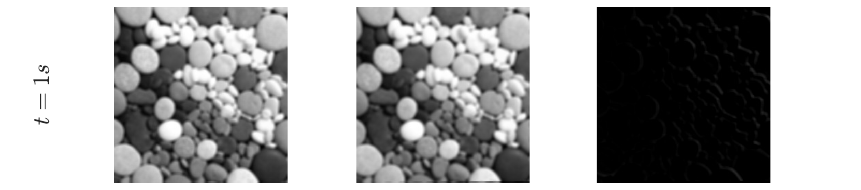}
\caption{Reference image $\mathring{\mathcal{I}}$, warped image $\mathcal{I}^e$ and difference image $\tilde{\mathcal{I}}$ at different time intervals.}
\label{fig:diff_image}
\end{figure}

\begin{figure}[h]
\centering 
%\scriptsize{Homography estimation error }\medskip
\includegraphics[scale=.4]{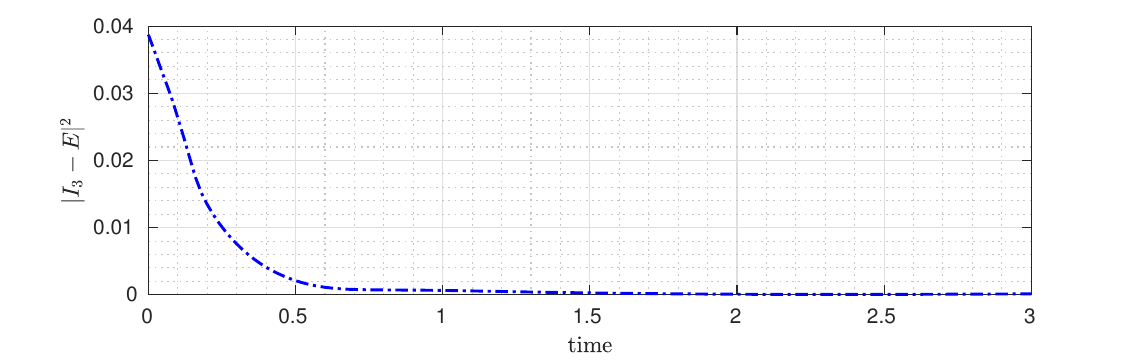} \\
\includegraphics[scale=.4]{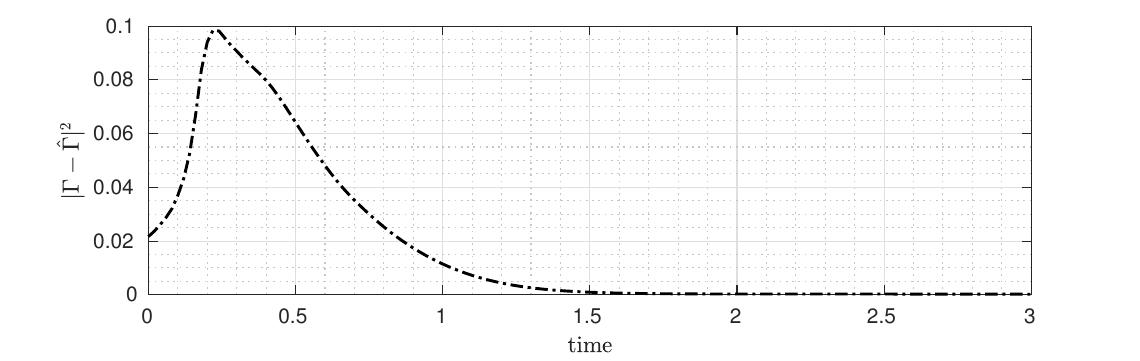} \\
\includegraphics[scale=.4]{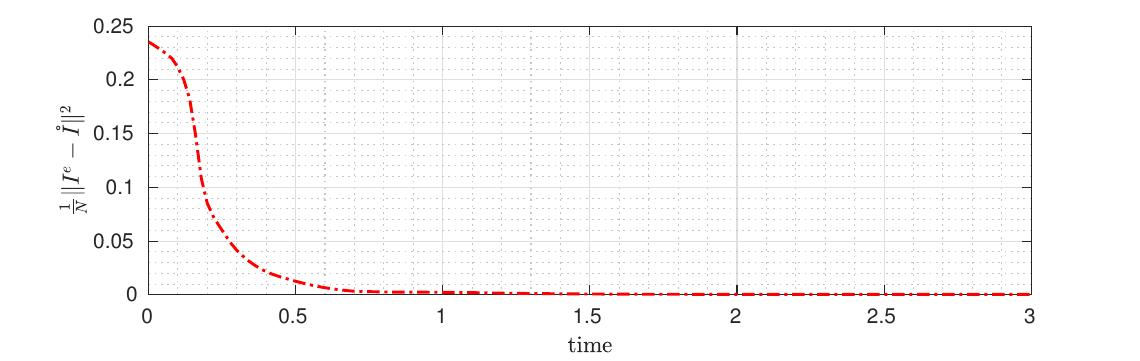}
\caption{Time evolution of the homography estimation error (blue), partial velocity estimation error (black) and image intensity error (red) of the second simulation.} 
\label{fig:errors2} 
\end{figure}

The first simulation results are shown in Figures \ref{fig:errors} and  \ref{fig:diff_image}.
Figure \ref{fig:errors} illustrates the estimation error $\epsilon_H = |I_3 - E |^2$ and the image intensity error between the reference and warped images divided by the number of pixels $\epsilon_I = \frac{1}{N}\|I^e - \mr{I} \|^2$.
Figure \ref{fig:diff_image} displays the warped image and the resulting difference image with respect to the reference at $t=0s$, $t=0.15s$ and $t=1s$. It clearly shows that the warped image converges towards the reference image with time.
The estimation errors $\epsilon_H$, $\epsilon_I$ and $\epsilon_\Gamma =  |\Gamma - \hat{\Gamma} |^2$ of the second simulation are illustrated in Figure \ref{fig:errors2}. 

The simulation results show clear convergence of the estimation errors to zero and validate the effective estimation performance of the proposed observers. 

\section{Conclusion}
In this paper, we introduced a nonlinear observer for dynamic homography estimation using a direct image error.
The proposed solution uses images as measurements to compute the homography between two given frames by exploiting an induced action of $\SL(3)$ on the space of image functions, and local asymptotic stability of the error is shown.
Additionally, an extension is proposed for velocity estimation that exploits the angular velocity measurement to estimate the remaining part of the group velocity when this is slowly time-varying.
The theoretical results were validated through simulation, where the image difference and homography estimation error are shown to rapidly converge to zero.

\section*{Acknowledgment}
This work has been supported by the French government, through the EUR DS4H Investments in the Future project managed by the National French Agency (ANR) with the reference number ANR-17-EURE-0004.%

\bibliographystyle{unsrtnat}
\bibliography{references}  

\end{document}